\documentclass{raa_twocolumn}            

\usepackage{graphicx,times}             
\usepackage{natbib}
\usepackage{amssymb,amsmath}
\bibpunct{(}{)}{;}{a}{}{,}
\usepackage[pagebackref=true]{hyperref}

\begin{document}

  \title{SVOM/ECLAIRs detection plane: main features and performance}

   \volnopage{Vol.0 (202x) No.0, 000--000}      
   \setcounter{page}{1}          

   \author{O. Godet 
      \inst{1,*}\footnotetext{$*$Corresponding Authors, these authors contributed equally to this work.}
   \and J.-L. Atteia
      \inst{1}
   \and R. Pons
      \inst{1}
    \and C. Amoros
      \inst{1}
     \and V. Waegebaert 
      \inst{1}
    \and K. Lacombe
      \inst{1}
    \and P. Guillemot
      \inst{4}  
    \and B. Arcier
    \inst{1}
     \and A. Bajat
    \inst{1}
    \and D. Barret
      \inst{1} 
    \and L. Bouchet
      \inst{1}
     \and J.-P. Dezalay
      \inst{1}
    \and O. Gevin
      \inst{3}
      \and S. Guillot
      \inst{1}
    \and O. Limousin
      \inst{2}
    \and S. Maestre
    \inst{1}
        \and S. Mate
    \inst{1}
    \and K. Mercier  
     \inst{4} 
    \and G. Nasser
    \inst{1}
    \and L. Perraud  
    \inst{4} 
    \and D. Rambaud
    \inst{1}
    \and P. Ramon
      \inst{1}
    \and N. Remoué
    \inst{1}
    \and M. Yassine 
    \inst{1}
 }

   \institute{IRAP, Université de Toulouse/CNRS/CNES, 9 avenue du colonel Roche, 31028 Toulouse, France; {\it ogodet@irap.omp.eu}\\
        \and
        Université Paris-Saclay, Université Paris Cité, CEA, CNRS, AIM, F-91191 Gif-sur-Yvette, France;\\
        \and
        Université Paris-Saclay, Institut de Recherche sur les lois Fondamentales de l'Univers, CEA, F-91191 Gif-sur-Yvette, France;
        \and
        CNES, 18 Avenue Edouard Belin, 31401 Toulouse cedex 9, France\\
\vs\no
   {\small Received 202x month day; accepted 202x month day}}

\abstract{ 
The detection plane of the high-energy transient camera ECLAIRs onboard {\it SVOM} is made of 200 XRDPIX detection modules, each consisting of a matrix of $8\times 4$ Schottky-type CdTe detectors hybridized with the low-noise and low-consumption ASIC IDeF-X. The 6400 detectors are operated at --20°C and reverse biased at --300\,V, reaching a low energy threshold of 4 keV. The $20\,\mu\mathrm{s}$ time resolution readout electronics works in photon counting mode classifying detected events as single or multiple events and measuring the deposited energy in real time. In this paper, we present the detection plane sub-systems and their main features. We also discuss its overall performances as measured both on ground and in-flight, showing compliance with the ECLAIRs science requirements.
}

   \authorrunning{O. Godet et al.}            
   \titlerunning{ECL detection plane}  

   \maketitle

%
%
\section{Introduction}           
\label{sect:intro}

ECLAIRs (hereafter ECL -- \citealt{Godet14, Godet25}) is the high-energy transient trigger camera onboard the Space-based multi-band Variable Object Monitor ({\it SVOM} mission launched successfully on June 22, 2024 (\citealt{Wei16, Cordier25}). 
SVOM is dedicated to study the high energy (HE) transient sky with a particular interest for Gamma-ray bursts (GRBs) signaling the catastrophic formation of stellar-mass black holes/neutron stars and the launch of powerful ultra-relativistic jets oriented towards Earth (\citealt{Zhang07}). {\it SVOM} embarks on an agile platform a suite of two wide field instruments (ECL and the Gamma-ray Monitor -- \citealt{GRM25}) to study the GRB prompt emission from 4 keV to 5 MeV and two narrow-field follow-up telescopes: the Micro-channel X-ray Telescope (MXT -- \citealt{MXT25}) and the Visible Telescope (VT -- \citealt{VT25}). Since {\it SVOM} is in a low Earth orbit (LEO) with an inclination of 29°, the satellite crosses the deep South Atlantic Anomaly (SAA) 5–6 times per day. 
Due to the SVOM near anti-solar pointing, Earth crosses the instrument field of view (FoV) every orbit.

ECL provides both the autonomous detection and first localization of new transients found within its 2 sr FoV thanks to its two trigger software (\citealt{Schanne25}). It is designed as a 4 -- 150 keV coded mask camera based on a 2-D coded mask (\citealt{Lachaud25}) placed at $\sim 46$ cm above a pixelated detection plane, both elements being enclosed by a passive graded shield made of lead and copper layers (\citealt{Godet25}).  
The present paper focuses on the core of the ECL camera i.e. the detection plane (DET) and its readout electronics (Electronique de Lecture Secteur -- ELS), both built by IRAP under the supervision of the French space agency (CNES). 

In Section 2, we describe the main parts of the DET and ELS. Section 3 gives a summary of the on-ground and in-flight performance results. Section 4 is devoted to the main conclusions.

  \begin{figure*}
        \centering
        \includegraphics[width=\linewidth]{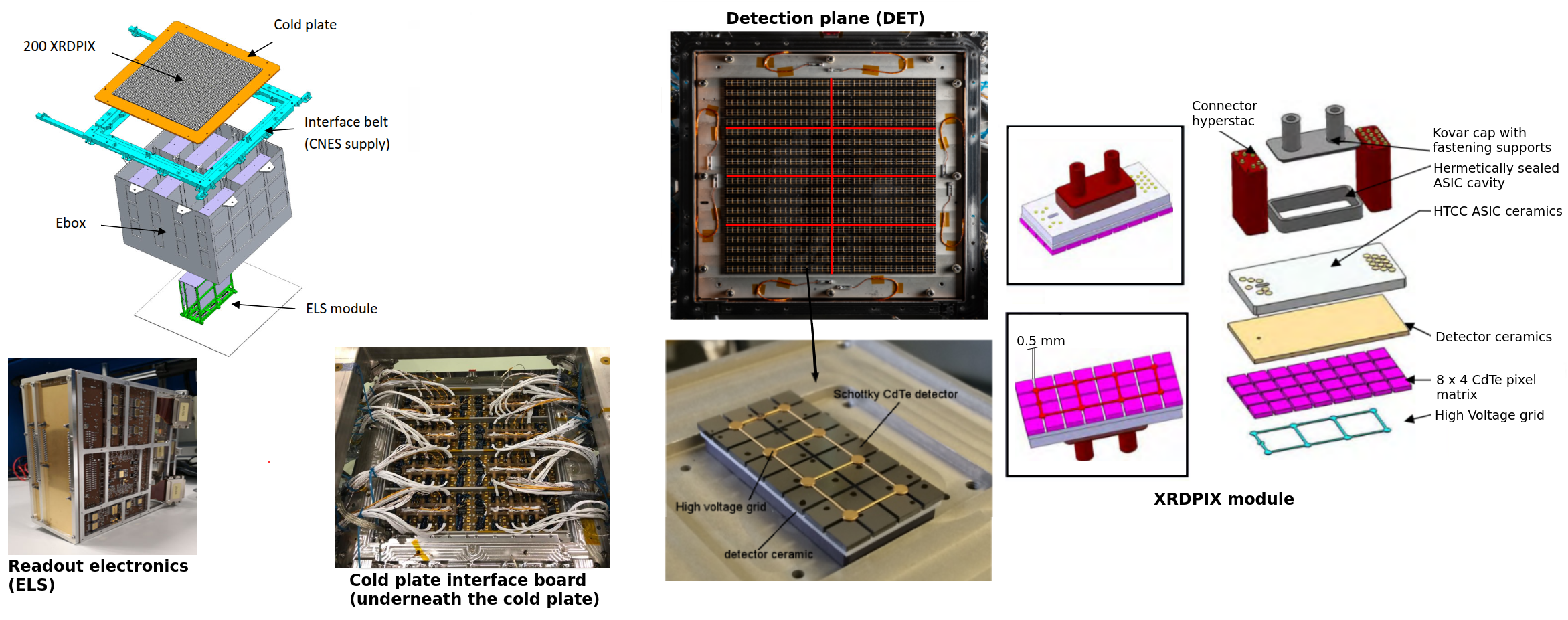}
        \caption{Pictures of the flight model of the detection plane made of $80 \times 80$ Schottky-type CdTe pixels and its readout electronics (ELS). The detectors are mounted on a cold plate as elementary detection modules called XRDPIX, each consisting of a $8 \times 4$ pixel matrix hybridized with the low noise and low power consumption ASIC IDeF-X from CEA. } 
        \label{fig:DPIX}
    \end{figure*}

\section{Description of the detection plane and readout electronics}
\label{sect:DPIX}


To enhance the ECL sensitivity to soft gamma-ray bursts and high-z GRBs (\citealt{Godet09}), ECL has a low energy threshold at 4 keV, compared to 14 keV for the Burst Alert Telescope onboard the Neil Gehrels {\it Swift} observatory (\citealt{Swift, BAT}). This science requirement drove the design of the whole instrument (\citealt{Lachaud25, Godet25}) and in particular the detection plane.  
 
The ECL detection plane is paved by 200 XRDPIX detection modules (\citealt{L13} -- see Section~\ref{sect:XRDPIX}), totaling $80 \times 80$ Schottky-type CdTe pixels, each of $4\times 4~\mathrm{mm}^2$ and 1\,mm thick (see Fig.\ref{fig:DPIX}). The detector cathode is made of a $\sim 250$\,nm-thick platinum layer (ohmic contact), while the anode's Schottky contact is made of a $300$\,nm-thick indium layer with a $\sim 50$\,nm-thick titanium layer in-between the CdTe and the indium anode. The DET dimensions are $36 \times 36$ cm$^2$ for a total geometrical surface of $\sim 950$ cm$^2$. 14400 detectors were purchased to Acrorad Co., Ltd. in Japan (\citealt{Funaki07}). 1 millimeter-thick Schottky CdTe detectors exhibit high absorption power in the 4 -- 150 keV band ($\sim 20\%$ at 150 keV) and low ($< 100$ pA) leakage current when reverse-biased at voltage values down to --600\,V and moderately cooled to approximately --20°C -- see \cite{Remoue10}. However, such detectors are known to suffer from the polarization effect when biased that degrades their spectroscopic and efficiency performances with time (\citealt{Cola09, Meuris11}). Fortunately, this effect is reversible when depolarizing the detectors for a few minutes. In orbit, the detectors are therefore depolarized when the satellite goes through the deep SAA (\citealt{Godet25}). 

During the development phase, all the detectors went through a series of measurements: leakage current measured at --20°C and +25°C and spectroscopic measures at +25°C using an $^{241}$Am source with detectors polarized at --600V. From this series of measures, we established several criteria to select detectors with the best performances (\citealt{Remoue10}). From the sample of 14400 purchased detectors, 8000 were selected to build the flight model (6400 pixels) and two spare sectors ($2\times 800$ pixels) that are currently used as a ground model at the ECL Instrument Centre\,\footnote{The EIC is the centre of expertise of the ECL instrument in charge of monitoring its health and the evolution of its performances, updating the calibration and onboard configuration files as well as the onboard software.} (EIC) located at Toulouse (France).         

The XRDPIX modules are mounted on a cold plate made in AlBeMet-162 (i.e. 62\% in beryllium and 36--40\% in aluminum). Underneath the cold plate, there is a series of constant conductance heat-pipes (CCHPs) used to cool down the detectors within their nominal temperature range with a mean value of --20°C (see \citealt{Godet25}). The detector temperature and the cold plate regulated temperature differ by 5°C.

Therefore, to reach a detector temperature of --20°C, the cold plate is regulated to a temperature of --25°C.      
For redundancy purposes, the DET is divided into 8 sectors, each made of 800 detectors (i.e. $5\times 5$ XRDPIX modules). Each sector is coupled to an ELS readout electronics, which are almost electrically independent from each other (see Section~\ref{sect:ELS}).

\subsection{XRDPIX modules}
\label{sect:XRDPIX}

Figure~\ref{fig:DPIX} (right) shows the anatomy of an XRDPIX module. It is made of a matrix of $4 \times 8$ Schottky CdTe detectors with a pitch of 4.5 mm stick on the anode side on an Al$_2$O$_3$ ceramic, hybridized to the low noise and low power consumption ASIC IDeF-X developed by CEA (\citealt{Gevin07, Gevin09}). The ASIC allows to read out the signals from the 32 detectors. The ASIC glued on a circuited ceramic is encapsulated in a kovar cavity hermetically sealed. Two fastening supports allow to fix the module onto the cold plate. A grid in kovar to polarize the detectors is glued on the detector cathode side. An epoxy glue rich in silver is used to stick together all the module components. Hyperstac connectors are used to transfer analog and digital signals to the ELS. 

For nominal in-flight operations, the detectors are cooled down to --20°C and reverse-biased at --300 V.

\subsection{Readout electronics}
\label{sect:ELS}

Each ELS is made of 4 electronic boards: 1) The pre-processor board uses the NanoXplore FPGA (Field Programmable Gate array) operating at 10\,MHz. The FPGA controls the readout process of the signals produced by the 25 XRDPIX modules of a sector; 2) The MUX-ADC board is used for multiplexing and analog-to-digital conversion of the signals coming from the sector modules. This board is controlled by the FPGA; 3) The LV-HVPS board generates all the voltages required by the different components of the ELS and the XRDPIX modules, including the ASIC; 4) The backplane board is required to interconnect the ELS boards.

{\bf Event readout} -- When an event is detected above the low-level discriminator (LLD), the event electronic signal is integrated for $t_\mathrm{peaking} = 2.6\,\mu\mathrm{s}$. After $t_\mathrm{freeze} = 3.2\,\mu\mathrm{s}$, the ELS FPGA starts the readout process and the detection module is no longer able to detect new events. At this stage, the FPGA encodes as a 32 bit frame the event detection time, its position on the DET and its energy depending on its type (see below). The event energy is encoded over a 12 bit dynamics and then rounded to 10 bits. Events that accumulate in channel 1023 are considered to be saturating events. 

The event time tagging is performed with a resolution of $20~\mu\mathrm{s}$ using a 9 bit counter synchronized on the time frame events (TFE) emitted every 10\,ms by the ECL data processing unit (Unité de Gestion et de Traitement Scientifique -- UGTS, \citealt{Godet25}). Using the pulse per second signal sent to the UGTS by the payload data processing unit, it is then possible to compute the event detection time synchronized to the satellite onboard time.  

The duration of the event readout process depends on the event type and represents a dead time for the module being read out. Thus, for a single event, the total dead time is around 59.5 $\mu$s, while for a double event on the same module, this time is around 69.4 $\mu$s (\citealt{Bajat18b}). The total time to encode the energy of one event is 16.7 $\mu$s. 
To mitigate for inter-module cross-talk, when pixels are triggered on a given module (M), the FPGA also freezes the adjacent module to its left (M$-$1). Consequently, any photons arriving at the module M$-$1 during the readout period of module M is lost, introducing additional dead time in module M$-$1 (see Section~\ref{deadtime}).

{\bf Event classification} --  The ELS classifies detected events as single or multiple events in order to discriminate photons depositing all their energy in one detector (single events) from multiple events induced by fluorescence/Compton interactions and particle showers hitting several detectors. Single events (SE) are further tagged into four adjustable energy bands in keV (EBANDS = [5 -- 8], [8 -- 20], [20 -- 50], [50 -- 120]). They are called SEs with bound energies (SEBs). Only SEBs are used by the onboard trigger software (\citealt{Schanne25}). Multiple events are detected at the XRDPIX and sector levels over a $t_\mathrm{freeze} = 3.2\,\mu\mathrm{s}$ or $\mathrm{ACD} = 1\,\mu\mathrm{s}$ coincidence time window, respectively. $t_\mathrm{freeze}$ and ACD are adjustable parameters. 
To reduce the readout induced dead time, the parameters \textsc{asic$_{-}$limit} and \textsc{sector$_{-}$limit} have been set to 2 so that the energy encoding is only performed for single and double events at XRDPIX and sector levels, respectively. These events carry information about fluorescence interactions within detectors that are useful for in-flight calibration activities (\citealt{Godet25}). For other multiple events, the ELS only records their detection time and either the number of pixels hit within a detection module (for \textsc{asic$_{-}$limit} $> 2$) or the number of modules hit over a sector (for \textsc{sector$_{-}$limit} $> 2$).  
On a sector (800 pixels), the ELS also count the global event counts per pixel prior to event classification over a 8\,s time frame (i.e. the TFE events emitted every 10\,ms each time for a different pixel of the sector).  
The ELS can produce up to $\sim 28000$ event frames s$^{-1}$.

\medskip

The ELS and ASIC parameters mentioned above as well as the value of the detector high voltage were fine-tuned prior to the launch through a series of measurements performed on several DET prototypes in order to optimize the DET performances and operability (\citealt{Nasser15, Bajat18c}). The main parameters defining the camera nominal working point are summarized in Table~\ref{tab:workingpoint}.     

\begin{table}
\begin{center}
\caption[]{Main parameters defining the camera nominal working point.}\label{tab:workingpoint}
 \begin{tabular}{ll}
  \hline\noalign{\smallskip}
Parameter &  Characteristic                 \\
  \hline\noalign{\smallskip}
Peaking time $t_\mathrm{peaking}$ & $2.6\,\mu\mathrm{s}$ \\
Time to freeze $t_\mathrm{freeze}$ & $3.2\,\mu\mathrm{s}$ \\
Coincidence time window ACD &  $1\,\mu\mathrm{s}$ \\
at sector level &  \\
\textsc{asic$_{-}$limit}* & 2 \\ 
\textsc{sector$_{-}$limit}* & 2 \\
Time resolution & $20\,\mu\mathrm{s}$\\

 \hline 
 \hline
   \noalign{\smallskip}
Detector high voltage  &   --300\,V                 \\
Detector temperature & --20°C \\
  \noalign{\smallskip}\hline
   \hline
   \noalign{\smallskip}
Energy bands for trigger software & 5 -- 8; 8 -- 20;  \\
 (keV)                            & 20 -- 50; 50 -- 120 \\
  \noalign{\smallskip}
   \hline
   * Energy encoded only for & \\
   single and double events. & \\
\end{tabular}
\end{center}
\end{table}

\section{DET performances}
\label{sect:perf}

\subsection{Spectral performances}
\label{sect:spec}

  \begin{figure}
   \centering
   \includegraphics[width=0.4\textwidth]{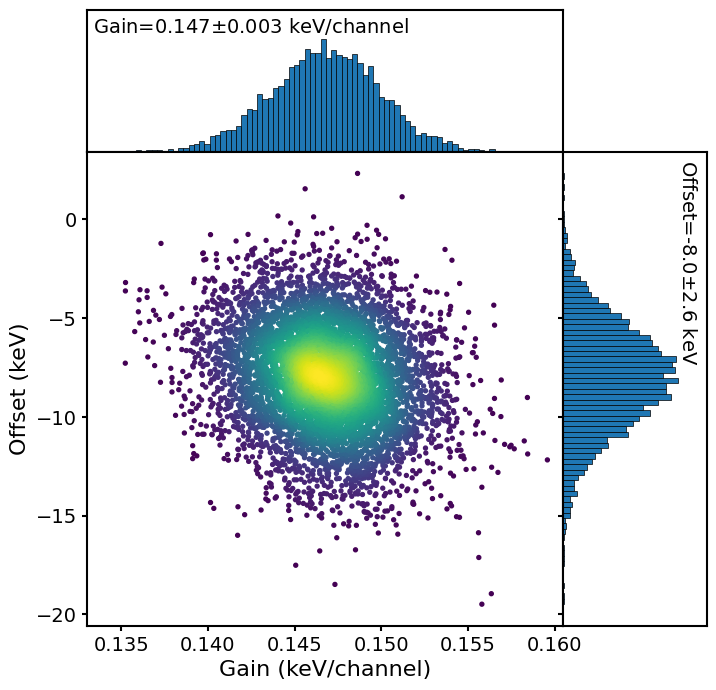}
   \caption{Distribution of the Gain and Offset coefficients for the 6400 pixels of the detection plane. Their medians and 1\,$\sigma$ uncertainties are indicated in the top and right panels. }
   \label{fig:gain_offset}
   \end{figure}

{\bf Energy scale} -- To calibrate the linear relationship between the encoded energy in channel $C_{\rm evt}$ and its reconstructed energy in keV $E_{\rm evt}$ for each DET pixel: $E_{\rm evt} = \mathrm{Gain} \times C_{\rm evt} + \mathrm{Offset}$, we measured $^{55}$Fe + $^{241}$Am spectral lines for an exposure time of several hours during the on-ground tests. Spectral line fitting from 5.9 keV up to 60 keV allowed us to derive the line centroid $C_\mathrm{line}$ and associated $1\,\sigma$ errors. We used Gaussian profiles for lines with an energy below 40 keV, while more sophisticated line profiles were needed at high energies to account for line asymmetry on the low-energy side due charge losses (\citealt{Bajat18a}). A linear fit from the $C_\mathrm{line}$ -- $E_\mathrm{line}$ datapoints with $E_\mathrm{line}$, the expected line energy in keV, allowed to compute the Gain and Offset coefficients for each DET pixel. Figure~\ref{fig:gain_offset} shows the distribution of the Gain--Offset coefficients. Note the very homogeneous distribution of the derived coefficients. The Pulse Invariant (PI) channel width is around 0.15 keV. 

  \begin{figure}
   \centering
   \includegraphics[width=0.4\textwidth]{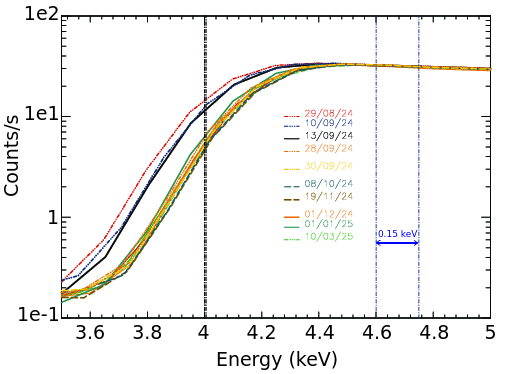}
   \includegraphics[width=0.4\textwidth]{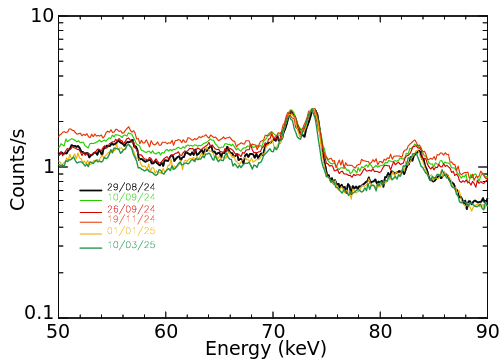}
   \caption{(Top) In-flight background spectra below 5 keV from August 29, 2024 to March 10, 2025. The two thin blue vertical lines indicate the size of a PI channel (0.15 keV).  
   (Bottom) In-flight background spectra in the 65 -- 100 keV energy range from August 29, 2024 to March 10, 2025. The spectra are normalized to the peak count rate of the Pb K$\alpha1$ line obtained on August 29, 2024.
   }
   \label{fig:SBNEscale_stability}
   \end{figure}

Figure~\ref{fig:SBNEscale_stability} (bottom panel) shows the Pb K$\alpha,\beta$ and Ta K$\alpha$ fluorescence lines on the instrument background spectra measured in-flight at different epochs up to the end of March 2025. To ease the comparison between epochs, the data were selected so that the Earth was not within the ECL FoV and for times when the satellite was far from the SAA regions. From this figure, the instrument energy scale appears to be stable over the selected time window since launch even if there is a shift of $\sim 0.2$ keV towards lower energies for the March 2025 spectrum. The gain/offset coefficients will be shortly updated. See also \cite{Godet25}. 

\medskip

{\bf Low energy threshold} --  We calibrated on-ground the linear relationship SBN\,\footnote{SBN stands for Seuil Bas Numérique (i.e. digital low-energy threshold).} -- $E_{\rm keV}$: $E_{\rm keV} = A \times \mathrm{SBN} + B$, using $^{55}$Fe $+$ $^{241}$Am spectra with 4 SBN values (27, 40, 50, 60) set for all detectors (\citealt{Godet22}). SBN is an integer from 0 to 62 allowing to choose the value of the ASIC LLD. SBN = 62 roughly corresponds to a low-energy threshold of 14 keV. The $E_{\rm keV}$ values associated at each SBN value was derived using a threshold method (here the threshold value was fixed at $3\times 10^{-3}$ counts s$^{-1}$). We then computed the $A$ and $B$ coefficients (expressed in keV) through a linear fit for each detector (\citealt{Godet22}). The average $A$ value is 0.24 keV i.e. 1.6 PI channels.  

   \begin{figure}
   \centering
   \includegraphics[width=0.4\textwidth]{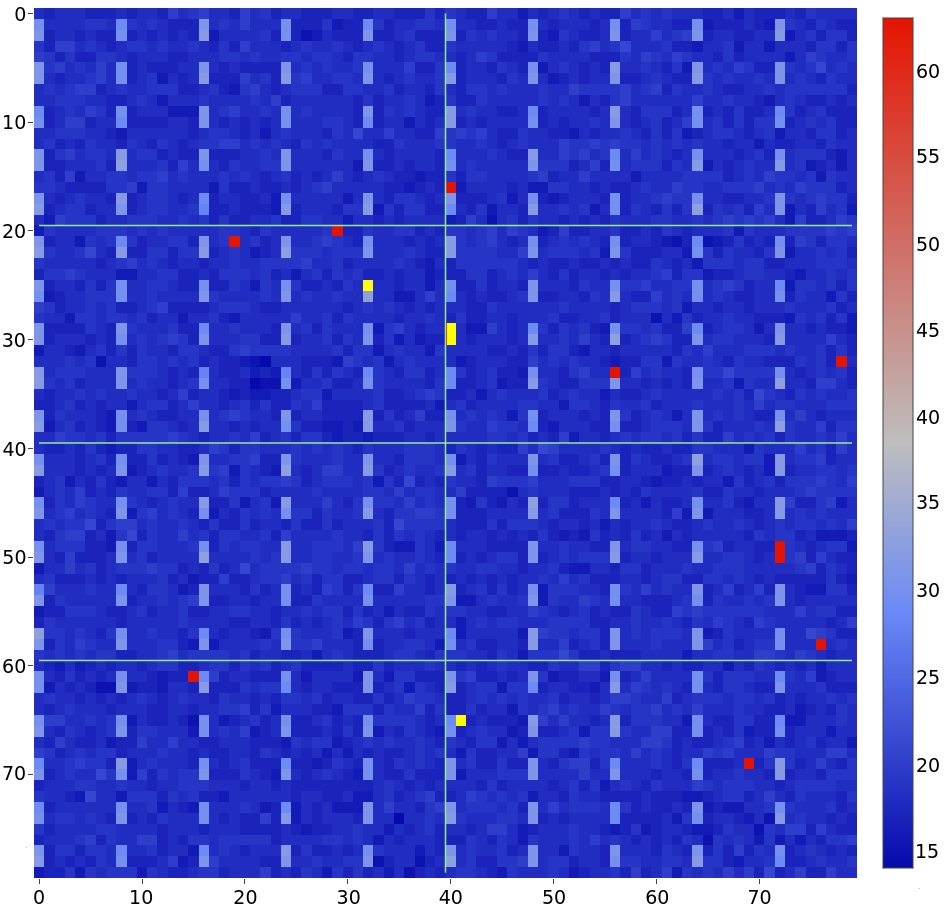}
   \caption{Map showing the SBN values for the 6400 pixels of the detection plane. In red, the 10 dead pixels identified prior to the launch with SBN = 63. In yellow, the 4 dead pixels (SBN = 63) added since then. In light blue, the pixels with a low-energy threshold set at $\sim 7$ keV to mitigate cross-talk issues. For the other pixels, their low-energy threshold has been set at $\sim 3.8$ keV.}
   \label{fig:threshold_inflight}
   \end{figure}

The low energy threshold was set at $E_\mathrm{thres}\sim 3.8$ keV for $93.5\%$ DET pixels, while 400 pixels have a low energy threshold set at $E_\mathrm{thres} \sim 7$ keV to mitigate cross-talk issues at the XRDPIX level (see Section~\ref{sect:dead}). Using the calibrated SBN -- $E_{\rm keV}$ relationship for each detector, we computed the SBN value as the closest integer to reach the required $E_\mathrm{thres}$ value (see Fig.~\ref{fig:threshold_inflight} and \citealt{Godet22}).   

  \begin{figure}
   \centering
   \includegraphics[width=0.4\textwidth]{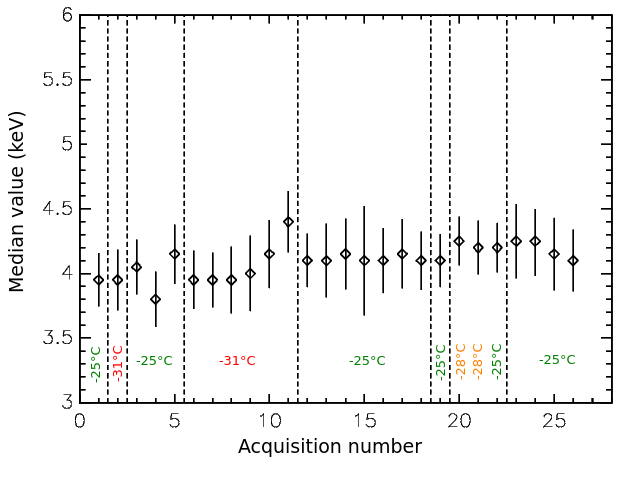}
   \caption{Evolution of the median value of the low-energy threshold for different radiator thermal conditions during on-ground thermal-vacuum tests on the whole instrument. The vertical bars correspond to the dispersion over the 6390 active pixels (on average $\pm 0.24$ keV i.e. $\pm 1$ increment of the SBN value). The temperatures indicated on the plot, separated by dashed lines, are the regulation temperatures of the cold plate.}
   \label{fig:SBN_stability}
   \end{figure}

On-ground thermal-vacuum tests of the whole ECL instrument to simulate different in-flight thermal environments for the thermal control system allowed to demonstrate the very good stability of the detector low-energy thresholds (see Fig.~\ref{fig:SBN_stability}). In-flight background data up to the end of March 2025 show that the detector low-energy thresholds are rather stable over time (see the top panel in Fig.~\ref{fig:SBNEscale_stability}). On September 27, 2024, the activation of heaters within the payload interface module increased the ELS box temperature by $\sim 7$°C. This resulted in an increase by around 0.15 keV of the detector low-energy thresholds as shown in  Fig.~\ref{fig:SBNEscale_stability} (top panel). This did not have any significant impact on the overall instrument spectroscopic performance, nor the trigger performance (\citealt{Schanne25}).

\medskip

\begin{figure}
        \centering
        \includegraphics[width=\linewidth]{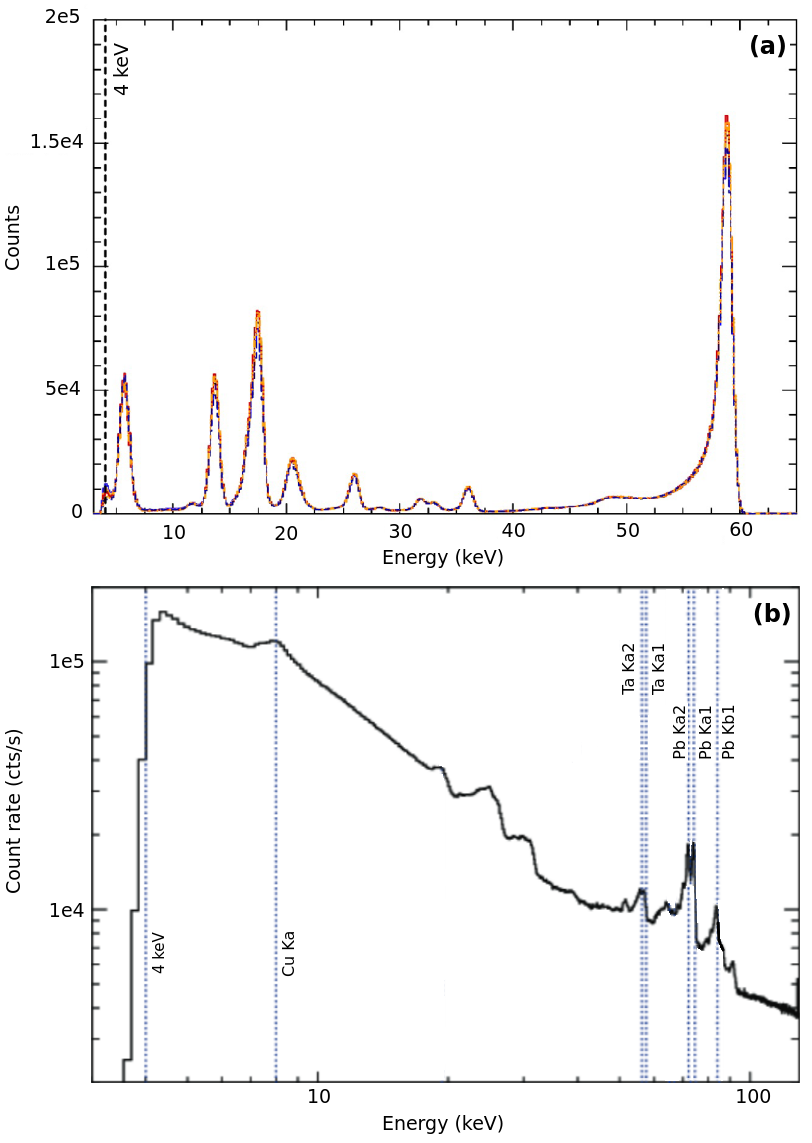}
        \caption{(a) Spectra for the 8 sectors of the DET measured during the on-ground tests. $^{55}$Fe (5.9 keV) and $^{241}$Am (10 -- 60 keV) radioactive sources were placed on-axis at 1\,m in front of the DET. (b) In-flight background spectrum when the Earth was outside the ECL FoV. The fluorescence lines from the mask and the shield are shown in the graph. } 
        \label{fig:bkg}
    \end{figure}

{\bf Spectral response} -- The top panel in Figure~\ref{fig:bkg} shows the superimposition of energy-calibrated spectra for the 8 DET sectors (i.e. stacking spectra from 800 pixels). These on-ground data were collected by illuminating the DET with radioactive sources ($^{55}$Fe $+$ $^{241}$Am) placed on-axis at 1\,m. The DET placed within a thermal-vacuum chamber was operated at its nominal working point. Note the overall homogeneity of the detector spectral response with energy. About 99.5\% of the detectors have a full width at half maximum (FWHM) less than 1.2 keV at 60 keV. 

At higher energies ($>$ 40 keV), the effects of charge losses become more and more important (\citealt{Bajat18b, Godet22}). During the on-ground tests, we also performed some on-axis spectral measures using $^{133}$Ba (30 -- 81 keV) and $^{57}$Co (7 -- 122 keV) -- see \cite{Godet22}. All these data have been used to calibrate our GEANT4 based numerical model of the DET spectral response. The ECL response matrix file (\citealt{Bajat18b, Bouchet26}) computed prior to the launch is currently used with success for spectral analysis of astrophysical sources (\citealt{Godet25}). 

The bottom panel in Figure~\ref{fig:bkg} shows an in-flight background spectrum when the Earth was not blocking the FoV and when the satellite was far from the SAA regions. The instrument fluorescence lines are clearly visible at low and high energies. These lines are used for the in-flight monitoring of detector energy scale (\citealt{Godet25}). The low-energy threshold of the whole camera did not change significantly after launch (see above).   

\subsection{Dead \& noisy pixels} 
\label{sect:dead}

A pixel is considered noisy when its count rate is much higher than the other pixels on the plane. We identified two types of noisy pixels from on-ground tests: 1) Those due to cross-talk from adjacent pixels; there are two such pixels on each XRDPIX module (i.e. 400 pixels on the DET). Additional counts appear either below the pixel low-energy threshold or as a peak energy below 7 keV. This pixel cross-talk creates multiple events on the XRDPIX module resulting in an artificial decrease in the number of single events. To prevent these effects, their low-energy threshold has been set up to $\sim 7$ keV instead of $\sim 4$ keV; 2) Pixel instability resulting in pixels becoming suddenly noisy over some time intervals before calming down. Most of the additional counts on the spectrum of these unstable pixels lie below 6 keV typically. 
In some rare cases observed during on-ground tests and in-flight, this unstable behavior can result in an entire XRDPIX module or a column of modules on a sector becoming noisy (\citealt{Godet22}). Investigation suggests that it is not related to the aging of the detectors and the electronics.

To prevent very noisy pixels to perturb the functioning of the onboard trigger software, we designed a dedicated algorithm implemented in the UGTS that autonomously disables (setting SBN = 63) at most one very noisy pixel over the DET, when its TFE count rates exceed 25 counts s$^{-1}$ (i.e. 200 counts over 8\,s) -- see \cite{Godet25}.  
A disabled pixel is no longer read out by the ELS. Each time the satellite exits the deep SAA, the disabled pixels are re-activated with their initial SBN value.

From the on-ground tests, 10 pixels over 6400 pixels were permanently disabled. Such pixels are called dead pixels. Their number is adjusted on ground at the EIC. Since launch, we identified and disabled 4 additional dead pixels resulting in a decrease in the effective area by $\sim 0.22$\%; which is acceptable (see Fig.~\ref{fig:threshold_inflight}). The number of in-flight automatically disabled very noisy pixels is just a few per day and is rather stable over time.

\subsection{Detector efficiency}
\label{sect:EA}

\begin{figure}
   \centering
   \includegraphics[width=0.45\textwidth]{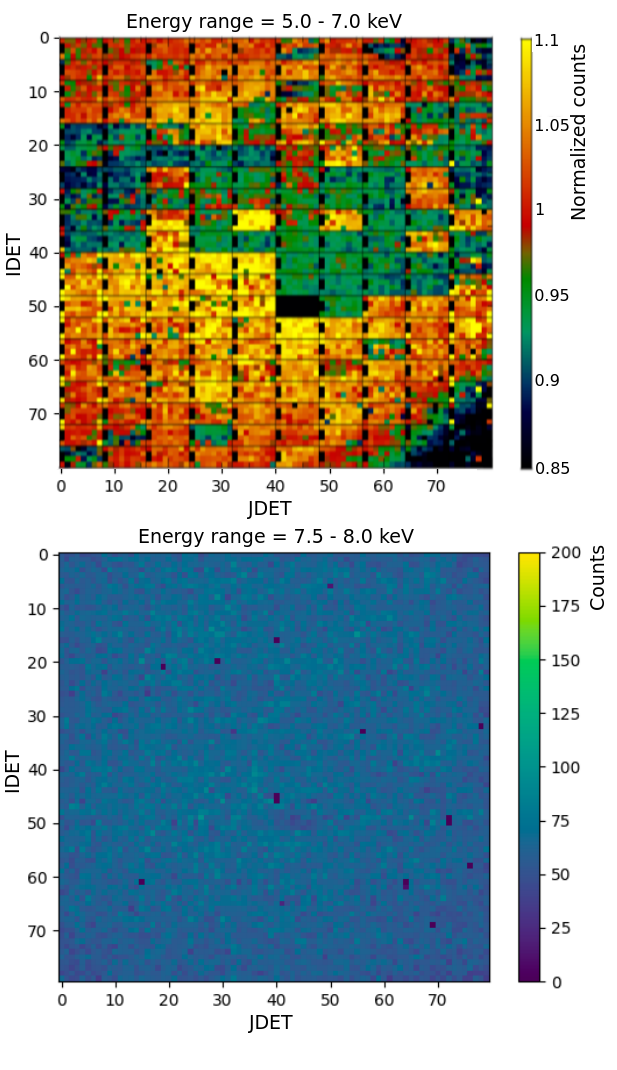}
   \caption{(Top) Count map in the 5 -- 7 keV band (around the $^{55}$Fe (5.9 keV) line) using data collected during the on-ground tests. The detector count rates are normalized to the median value of the count rates on the plane in the 5 -- 7 keV band.  
   The pixels in black have a 7 keV low-energy threshold. The module in black had an energy scale shifted downwards by almost 1 keV. (Bottom) In-flight background count map recorded on July 09, 2024 in the 7.5 -- 8 keV band. Note the good uniformity of the detector counts over the DET. Pixels in black are dead and disabled pixels.}
   \label{fig:Efficacity}
   \end{figure}

Figure~\ref{fig:Efficacity} shows count maps measured: 1) For the $^{55}$Fe 5.9\,keV line energy during on-ground tests (top panel). Note the 2-D curvature is due to the use of a point-like radioactive source placed at 1\,m from the DET. Some modules appear to count less (by $\sim 12\%$ at 5.9 keV) below 10 keV, while the detectors with a 7 keV energy threshold show no counts as expected; 2) For in-flight background data in the 7.5 -- 8 keV band when the Earth was not in the FoV and when the satellite was far from the SAA regions (bottom panel). Note the very good uniformity {\bf($<$ 3\%)} of the count rates over the plane in the bottom image.

The detectors that count less at low energies (shown in green in Fig.~\ref{fig:Efficacity}, top panel) come from a batch purchased in 2016, while the others correspond to detectors purchased in 2008. Investigation seems to point towards differences in the TeO$_2$ and/or Pt layers on top the detectors between the two batches. To correct all these detector efficiency discrepancies, \cite{Xie24} computed a detector efficiency table for different energy bands that is used onboard by the trigger software to correct the shadowgrams (\citealt{Schanne25}).

\subsection{Dead time} 
\label{deadtime}

\begin{figure}
   \centering
   \includegraphics[width=0.4\textwidth]{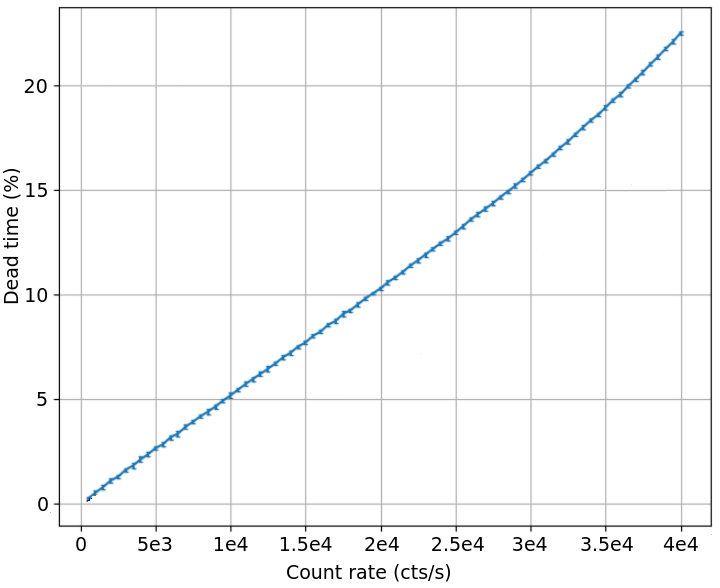}
   \caption{Evolution of dead time (i.e. the single event loss fraction) versus the incident count rate on a sector. The simulation has been performed at the sector level assuming that the 25 XRPDIX modules are illuminated in a uniform way,  ACD $= 1\,\mu$s, \textsc{asic$_{-}$limit} = 2 and \textsc{sector$_{-}$limit} = 2. }
   \label{fig:dead_time}
   \end{figure}

Dead time has to be computed at the sector level given the way the electronics work (see Section~\ref{sect:ELS} and \citealt{Bajat18b}). Figure~\ref{fig:dead_time} shows the fraction of lost single events versus the incident count rate assuming a uniform illumination of a sector. The ECL single event in-flight background count rate without the Earth blocking the FoV was estimated to be around 3000 counts s$^{-1}$ (i.e. 375 counts s$^{-1}$ sector$^{-1}$) in the 4 -- 120 keV band; which is in good agreement with what has been measured from in-flight data (\citealt{Godet25}). The single event count rate from the Crab source observed on-axis is around 540 counts s$^{-1}$ (i.e. 68 counts s$^{-1}$ sector$^{-1}$) in the 4 -- 150 keV band. In this case, the dead time is negligible. For an incident count rate of 12500 counts s$^{-1}$ sector$^{-1}$ (i.e. $10^5$ counts s$^{-1}$ on the whole plane), the event loss fraction is around 7\%. On-ground measures using radioactive sources placed at different distances in front of the DET allowed to validate our dead time model (\citealt{Bajat18b}). 

\section{Conclusion}

The detection plane of the ECL instrument overall displays excellent performances from on-ground and in-flight measures with only 14 dead pixels. The DET low-energy threshold was well set on-ground around $\sim 4$ keV for most pixels (93.5\%) and it is rather stable in orbit (varying within 0.15 keV i.e. one channel). 
Since launch, in-flight energy scale calibration was performed once to correct the Gain and Offset values of four XRDPIX modules (\citealt{Godet25}), and we adjusted the SBN values of 6 pixels (of which 4 were dead pixels with SBN = 63). We do not see any significant evidence for detector aging. 
Thanks to thorough and careful on-ground calibration tests, this allowed us to rapidly fine tune the trigger parameters as well as the SAA contours so that ECL started nominal operations as early as December 2024.

%

%

\begin{acknowledgements}
The Space-based multi-band astronomical Variable Objects Monitor (SVOM) is a joint Chinese-French mission led by the Chinese National Space Administration (CNSA), the French Space Agency (CNES), and the Chinese Academy of Sciences (CAS). We gratefully acknowledge the unwavering support of NSSC, IAMCAS, XIOPM, NAOC, IHEP, CNES, CEA, and CNRS. We thank all the persons (sub-contractors, students, engineers, etc.) who have also worked hard during more than 15 years to build a fantastic instrument. This paper is dedicated to Pierre Mandrou, a missed friend and colleague.

\end{acknowledgements}




\label{lastpage}

\end{document}